\documentclass[aps,prd,amssymb,eqsecnum,showpacs]{revtex4}

\begin{document}

\title{Worldtube conservation laws for the null-timelike evolution problem}

\author{Jeffrey Winicour${}^{1,2}$
       }
\affiliation{
${}^{1}$ Department of Physics and Astronomy \\
         University of Pittsburgh, Pittsburgh, PA 15260, USA \\
${}^{2}$ Max-Planck-Institut f\" ur
         Gravitationsphysik, Albert-Einstein-Institut, \\
	 14476 Golm, Germany \\
	 email: winicour@pitt.edu
	 }

\begin{abstract}

I treat the worldtube constraints which arise in  the null-timelike initial-boundary
value problem for the Bondi-Sachs formulation of
Einstein's equations. Boundary data on a worldtube and
initial data on an outgoing null hypersurface determine the exterior spacetime
by integration along the outgoing null geodsics. The worldtube constraints are a set
of conservation laws which impose conditions on the integration constants. I show
how these constraints lead to a well-posed initial value problem governing the
extrinsic curvature of the worldtube, whose components are related to the
integration constants.  Possible applications to gravitational waveform
extraction and to the well-posedness of the null-timelike initial-boundary value
problem are discussed.

\end{abstract} 

keywords: conservation laws, gravitational waves, boundary conditions 

\maketitle

\section{Introduction}

It is extremely gratifying to contribute this article in appreciation of Josh
Goldberg's friendship and guidance, especially because this opportunity to
recall Josh's early work has led me to an interesting approach to a current
problem. Josh and I first overlapped in 1963 when he came to Syracuse University
as a new Professor. At that time I was very busy finishing my PhD thesis and our
interaction was fortuitous in more ways than one. Not only did we share the same
research interests but the next year, when I was looking for my first position,
Josh recommended me to the Aerospace Research Laboratory at
Wright Patterson Air Force Base, where in his prior position he had organized
a general relativity research group which included such budding young
relativists as Roy Kerr. It was curious to me that the
Air Force sponsored research in a topic with no apparent military relevance.
Several years later, in a cost-cutting measure, Congress also came to this
curious realization and the general relativity group was disbanded. However,
during the intervening years the lab was a Camelot for basic research, much to
the credit of Josh's legacy. It was there that I came to meet and work with the
organizers of this volume, David Robinson and Ed Glass, two PhD students of Josh
who came to the Lab on  National Academy of Science postdoctoral fellowships.

Much of Josh's research centered around conservation laws and null
hypersurfaces~\cite{josh1,josh2,josh3}.
In the 1960's, these two topics were the focus of some of the most exciting
results in gravitational theory. They came together in Bondi's~\cite{bondi} and
Sachs's~\cite{sachs}   treatment of Einstein's equations via the characteristic
initial value problem, which led to the formulation of the Bondi mass $M_B$ and
news function $N$ as the quantities of prime physical importance for an isolated
system. The conservation law
\begin{equation}
          \frac {dM_B}{du} = - \oint |N|^2 \sin\theta d\theta d\phi,
          \label{eq:bmassl}
\end{equation} 
which relates the retarded time derivative of the Bondi mass to the integral of
the news function over the sphere at future null infinity, was the conclusive
theoretical evidence that ended any serious debate over whether gravitational
waves carry energy from a system. The conformal compactification of future null
infinity into the boundary ${\cal I}^+$ of an asymptotically flat spacetime put
this mass loss relation into a well-defined geometrical setting~\cite{Penrose}.
One of Josh's little known contributions to these results was the grant support
of Bondi's King College group, which he arranged through the Lab.

The Bondi mass loss equation is a result of certain constraints that arise in
the characteristic formulation of Einstein's equations.  The conservation law
(\ref{eq:bmassl}) is obtained by applying these constraints at ${\cal I}^+$. In
this paper, I  will discuss the content of these constraints when applied on a
timelike worldtube of finite size but surrounding the matter sources of the
gravitational field, as arises in the null-timelike initial-boundary value problem for
the gravitational field. There is much overlap between my results presented here
and Josh's work~\cite{josh3} which supplies the basic ideas for
interpreting these constraints as conservation laws. Josh applied these
conservation laws to the theory of equations of motion  for isolated
systems being developed by Newman and his students~\cite{posadas,lind}. Here I
apply these conservation laws to a problem in  numerical relativity that did not
exist at that time.  It is fitting, when I had a chance to talk to Josh at
a recent meeting on mathematical relativity, that he commented (essentially)
``these are the same problems that we tried to solve in the 1960's''. That is
true. The importance of those problems was recognized back then and only
recently have many of them been elevated to a high level status by the
mathematicians; and their importance to numerical simulation has also been
recognized. But they are the same problems!

In the Cauchy problem, initial data on a spacelike hypersurface ${\cal S}_0$ are
extended to a solution in the domain of dependence ${\cal D}({\cal S}_0)$ (which
consists of those points whose past directed characteristics all intersect
${\cal S}_0$).  In the initial-boundary value problem (IBVP), data on  a
timelike boundary ${\cal T}$  transverse to ${\cal S}_0$ is used to further
extend the solution to the domain of dependence
${\cal D}({\cal S}_0 \cup {\cal T})$.

The IBVP for Einstein's equations only recently received widespread attention
due to its importance to numerical relativity~\cite{stewart}, where the introduction
of a finite {\em artificial} outer boundary is standard practice. It is essential
for numerical evolution that the underlying analytic problem be well-posed. i.e.
that the solution depend continuously  on the data so that it is not
destabilized by numerical error. The first well-posed IBVP was achieved for a
formulation based upon a tetrad, connection and curvature as evolution
fields~\cite{friednag} and subsequently for the harmonic formulation of
Einstein's equations~\cite{krwi}.  However, much of the work in numerical relativity
is based upon other formulations where the well-posedness of the IBVP remains an
unresolved issue, incuding the null-timelike formulation. The
properties of the worldtube constraints treated here is important for a clearer
understanding of this characteristic IBVP.

I begin with a short review of the null-timelike IBVP in Sec.~\ref{sec:prob}. In
the traditional approach to the Cauchy problem for Einstein's equations, initial
data on a spacelike hypersurface with unit timelike normal $n_\mu$ are
formulated in a purely 3-dimensional form in terms of the intrinsic metric
$h_{\mu\nu}$ and extrinsic curvature $k_{\mu\nu}$ of the initial Cauchy
hypersurface. The components $G^{\mu\nu}n_\mu$ of the Einstein tensor contain
only first time-derivatives of the metric so that they constrain this
initial data, i.e. the Hamiltonian and momentum constraints
\begin{equation}
0={}^{(3)}R +(k^\mu_\mu)^2
      -k_{\mu\nu}k^{\mu\nu} \quad (= 2 G_{\mu\nu}n^\mu n^\nu)
      \label{eq:ham}
\end{equation}
and
\begin{equation}
 0 ={}^{(3)}\nabla_\mu (
  k^\mu_\nu - \delta^\mu_\nu k^\rho_\rho)\quad(= h_\nu^\mu 
    G_{\mu\rho} n^\rho),
    \label{eq:mom}
\end{equation}    
where ${}^{(3)} \nabla_\mu$ is the covariant derivative and ${}^{(3)}R$ is the
curvature scalar associated with $h_{\mu\nu}$. Subject to these constraints, the
Cauchy data determine a solution of Einstein's equations which is unique up to a
diffeomorphism (cf.~\cite{hawkel}).

These constraints are elliptic partial differential equations which couple their
solution with data on the boundary of the initial Cauchy hypersurface. Once
these constraints have been satisfied initially, the dynamical equations used in
a consistent hyperbolic reduction of Einstein's equations ensure that they continue
to be satisfied in time in the domain of dependence of the Cauchy problem.
Boundary conditions must then be formulated to preserve these constraints in the
larger domain of dependence  of the IBVP problem. That is the strategy which
lies behind the constraint-free Cauchy evolution schemes used in numerical
relativity. In characteristic versions of the IBVP, data are given on an initial
null hypersurface $u=const.$ This problem has the peculiar feature that the
normal co-vector $\nabla_\mu u$ is tangent to the hypersurface when interpreted
as a vector $k^\nu = g^{\mu\nu} \nabla_\mu u$ by the standard technique of
``raising indices'' with the metric. As a result, the corresponding Hamiltonian
and momentum constraints reduce to propagation laws consisting of ordinary
differential equations (ODE's) along the characteristics of the null
hypersurface. In the null-timelike version of the characteristic IBVP
the worldtube integration constants for these ODE's
uniquely determine the exterior spacetime. In Sec.~\ref{sec:prob} we review
these ODE's and their integration constants.

\section{The worldtube-nullcone problem}
\label{sec:prob}

\subsection{The null cone formalism}

We use coordinates based upon a family of outgoing null hypersurfaces. We let
$u$ label these hypersurfaces, $x^A$ ($A=2,3$) be labels for the null rays and
$r$ be a surface area coordinate. In the resulting $x^\alpha=(u,r,x^A)$
coordinates, the metric takes the Bondi-Sachs form~\cite{bondi,sachs}
\begin{equation}
   ds^2=-\left(e^{2\beta}{V \over r} -r^2h_{AB}U^AU^B\right)du^2
        -2e^{2\beta}dudr -2r^2 h_{AB}U^Bdudx^A +r^2h_{AB}dx^Adx^B,
   \label{eq:bmet}
\end{equation}
where  $det(h_{AB})=det(q_{AB})=q$, with $q_{AB}$ a unit sphere metric. The
contravariant components are 
\begin{eqnarray}
 g^{rr} &=& e^{-2\beta} \frac{V}{r} 
\label{eq:VH} \\
 g^{rA} &= &-e^{-2\beta} U^A \\
 g^{ru} &= & -e^{-2\beta}  \\
 g^{AB}&=&r^{-2} h^{AB} ,
\label{eq:rl}
\end{eqnarray}
where $h^{AB}h_{BC}=\delta^A_C$.

The Einstein equations $G_{\mu\nu}=0$ decompose into hypersurface equations,
evolution equations and conservation laws. We express these equations following
the formalism in~\cite{newt,nullinf}. The hypersurface equations correspond to
the components $G_\mu^\nu \nabla_\nu u=0$ and take the specific form
\begin{eqnarray}
   \beta_{,r} &=& \frac{1}{16}rh^{AC}h^{BD}h_{AB,r}h_{CD,r}
  \label{eq:beta} \\
    (r^4e^{-2\beta}h_{AB}U^B_{,r})_{,r}  &=&
   2r^4 \left(r^{-2}\beta_{,A}\right)_{,r}
     -r^2 h^{BC}D_{C}h_{AB,r}
   \label{eq:u} \\
   2e^{-2\beta}V_{,r} &=& {\cal R} - 2 D^{A} D_{A} \beta
   -2 D^{A}\beta D_{A}\beta + r^{-2} e^{-2\beta} D_{A}(r^4U^A)_{,r}
   -\frac{1}{2}r^4e^{-4\beta}h_{AB}U^A_{,r}U^B_{,r},
   \label{eq:v}
\end{eqnarray}
where $D_A$ is the covariant derivative and ${\cal R}$ the curvature scalar of
the 2-metric $h_{AB}$.

The evolution equations can be picked out by introducing a complex polarization
dyad $m^\mu$ satisfying $m^\mu \nabla_\mu u= 0$ which points in the
angular direction with components $m^\mu=(0,0,m^A)$ satisfying  $m^{(A} \bar
m^{B)}=h^{AB}$, so that $h_{AB}m^A \bar m^B =2$, which determines $m^A$ up to the
phase freedom $m^A \rightarrow e^{i \eta} m^A$. The evolution equations
correspond to the components $m^\mu m^\nu G_{\mu\nu}=0$ and take the form
\begin{eqnarray}
      m^A m^B\bigg  \{r (rh_{AB,u})_{,r}  &-& \frac{1}{2}( rV h_{AB,r})_{,r}
          -2e^{2\beta} (D_A D_B \beta
      + D_A \beta D_B \beta) \nonumber \\
       &+& h_{C(A} D_{B)}(r^2U^C)_{,r}
        -\frac{1}{2} r^4 e^{-2\beta}h_{AC}h_{BD} U^C_{,r} U^D_{,r} 
        +\frac{1}{2} r^2  h_{AB,r} D_C U^C \nonumber \\
    &+&r^2 U^C D_C h_{AB,r} 
       -D^C U_{(B}h_{A)C,r} +D_{(B}U^C h_{A)C,r}\bigg  \} =0.
\end{eqnarray}

If we introduce the auxiliary variable $Q_A =r^2 e^{-2\beta} h_{AB} U^B_{,r}$,
the system of hypersurface and evolution equations can be cast into a
hierarchy of first order ODE's,
\begin{eqnarray}
   \beta_{,r} &=&{\cal N}_\beta(h_{CD})
    \label{eq:nbeta}\\
   (r^2 Q_A)_{,r}  &=&{\cal N}_Q(h_{CD},\beta)
   \label{eq:nQ}\\
   U^A_{,r}&=& {\cal N}_U(h_{CD},\beta,Q_C)
   \label{eq:nU} \\
    V_{,r} &=&{\cal N}_V(h_{CD},\beta,Q_C,U^C) 
   \label{eq:nv} \\
    m^A m^B  (rh_{AB,u})_{,r} &=& {\cal N}_h(h_{CD},\beta,Q_C,U^C,V),
\end{eqnarray}
where the ${\cal N}$-terms on the right hand side can be calculated from the
values of their arguments on a given $u=const.$ null hypersurface. Each
${\cal N}$-term only depends upon previous members of the hierarchy in the order
$(h_{CD},\beta,Q_C,U^C,V)$. Thus, given $h_{AB}$ on the null hypersurface
$u=const.$, these equations can be integrated radially, in sequential order,
to determine $\beta$, $Q_A$, $U^A$, $V$ and $m^A m^B h_{AB,u}$ on the
hypersurface in terms of integration constants on an inner worldtube. For an
inner worldtube given by $r=R(u,x^A)$, the necessary integration constants are
\begin{equation}
   \beta|_R\, , \quad Q_A|_R \, ,  \quad U^A|_R \, , \quad  V|_R \, ,
    \quad m^A m^B h_{AB,u}|_R.
      \label{eq:intc}
\end{equation}
In addition, the location of the worldtube specified by $R(u,x^A)$ is another
essential part of the data.

\section{The boundary constraints  as conservation laws}
\label{sec:cons}

The components of Einstein's equations independent of the hypersurface and evolution
equations are the conservation conditions (called supplementary conditions by
Bondi and Sachs)
\begin{eqnarray}
        h^{AB}G_{AB}&=&0 \label{eq:triv} \\
        G_A^r&=&0 \label{eq:consu}\\
        G_u^r&=&0. \label{eq:consA}
\end{eqnarray}
As was shown by Bondi and Sachs, the Bianchi identity 
\begin{equation}
    \nabla_\mu G^\mu_\nu=\frac {1}{\sqrt{-g}} (\sqrt{-g}  G^\mu_\nu)_{,\mu}
          +\frac{1}{2} g^{\rho\sigma}_{,\nu} G_{\rho\sigma}=0
\end{equation}    
implies that these equations need only be satisfied on a worldtube $r=R(u,x^A)$.
When the hypersurface and evolution equations are satisfied, the Bianchi
identity for $\nu=r$ reduces to  $h^{AB}G_{AB}=0$ so that (\ref{eq:triv})
becomes trivially satisfied. (Here it is necessary that the worldtube have
nonvanishing expansion so that the areal radius $r$ is a non-singular
coordinate.) The Bianchi identity for $\nu=A$ then reduces to
\begin{equation}
          (r^2 G_A^r)_{,r} =0,
\end{equation}
so that $G_A^r=0$ if it is set to zero at $r=R(u,x^A)$. When that is the case, 
the Bianchi identity for $\nu=u$ then reduces to
\begin{equation}
          (r^2 G_u^r)_{,r} =0,
\end{equation}
so that $G_u^r=0$ also vanishes if it vanishes for $r=R(u,x^A)$.

As a result, the conservation conditions  can be replaced by the condition 
that the Einstein tensor satisfy
\begin{equation}
    \xi^{\mu}G_{\mu}^{\nu}\nabla_{\nu}r=0,
    \label{eq:Gcon}
\end{equation}
where $\xi^{\mu}$ is any vector field tangent to the worldtube.  This allows
these conditions to be interpreted as flux conservation laws for the
$\xi$-momentum contained in the worldtube~\cite{tam}. The unit norma
to the worldtube $N_\mu$ lies in the direction $\nabla_\mu(r-R(u,x^A))$, i.e. 
\begin{equation}
    N_\mu=\eta \nabla_\mu(r-R(u,x^A))
\end{equation}
where $\eta$ is a normalization constant. Since we we are assuming that the
hypersurface equations are satisfied, we can replace (\ref{eq:Gcon}) by 
\begin{equation}
    \xi^{\mu}G_{\mu}^{\nu}N_\nu=0.
    \label{eq:Gconr}
\end{equation}
These are the boundary analogue of the momentum constraints (\ref{eq:mom})
for the Cauchy problem. In a treatment of a timelike boundary for the
Cauchy problem, it has been pointed out~\cite{stewart} that the Cauchy momentum
constraints (\ref{eq:mom}) must be enforced on the boundary. Here, in the
characteristic initial-boundary value problem, it is the {\em boundary-momentum
constraints} (\ref{eq:Gconr}) which must be enforced.

Since $\xi^\mu N_\mu=0$, we can further replace (\ref{eq:Gconr}) by the
condition on the Ricci tensor
\begin{equation}
    \xi^{\mu}R_{\mu}^{\nu}N_{\nu}=0.
    \label{eq:Rcon}
\end{equation}
The Ricci identity
\begin{equation}
    \xi^{\mu}R_{\mu}^{\nu}= \nabla_\mu \nabla^{(\nu}\xi^{\mu)}
       + \nabla_\mu \nabla^{[\nu}\xi^{\mu]} -\nabla^\nu \nabla_\mu \xi^\mu
       \label{eq:xicon}
\end{equation}
then gives rise to the strict Komar conservation law
\begin{equation}
       \nabla_\mu \nabla^{[\nu}\xi^{\mu]} =0
\end{equation}
when $\xi^\mu$ is a Killing vector corresponding to an exact symmetry. More
generally,
(\ref{eq:xicon}) gives rise to the flux conservation law
\begin{equation}
           P_\xi(u_2)-P_\xi(u_1) = \int_{u_1}^{u_2} dS_\nu \{
     \nabla^\nu \nabla_\mu \xi^\mu - \nabla_\mu \nabla^{(\nu}\xi^{\mu)} \}
\end{equation}
where
\begin{equation}
           P_\xi =\oint dS_{\mu\nu} \nabla^{[\nu}\xi^{\mu]} 
\end{equation}
and $dS_{\mu\nu}$ and $dS_\nu$ are, respectively, the appropriate surface and
3-volume elements on the worldtube. For the limiting case when $R\rightarrow
\infty$, these flux conservation laws govern the energy-momentum, angular
momentum and supermomentum corresponding to the generators of the
Bondi-Metzner-Sachs asymptotic symmetry group~\cite{tam}. For an asymptotic time
translation, they give rise to the Bondi mass loss relation (\ref{eq:bmassl}).

Josh applied these conservation laws to a new treatment of equations of
motions in general relativity~\cite{josh3}.  Here I pursue a different application to
the mathematical basis of the worldtube constraints for the null-timelike IBVP.
For this purpose it is useful to rewrite the conservation laws (\ref{eq:Rcon}) in
terms of the intrinsic metric and extrinsic curvature of the worldtube.

The intrinsic metric of a worldtube embedded in the spacetime with  unit
spacelike normal $N_\mu$ is 
\begin{equation}
   H_{\mu\nu}=g_{\mu\nu}-N_\mu N_\nu
\end{equation}
and its extrinsic curvature is 
\begin{equation}
        K_{\mu\nu}=H_\mu^\rho \nabla_\rho N_\nu.
\end{equation}
We then have the worldtube analogue of (\ref{eq:mom}) for 
the Cauchy problem,
\begin{equation}
 0 ={\cal D}_\mu (
  K^\mu_\nu - \delta^\mu_\nu K^\rho_\rho)\quad(= H_\nu^\mu 
    G_{\mu\rho} N^\rho),
    \label{eq:bmom}
\end{equation}    
where ${\cal D}_\mu$ is the covariant derivative associated with $H_{\mu\nu}$.
These are equivalent to the conservation conditions (\ref{eq:Rcon}) and allow
the conserved quantities to be expressed in terms of the extrinsic curvature of
the boundary. For any vector field $\xi^\mu$ tangent to the worldtube,
(\ref{eq:bmom}) implies
\begin{equation}
   {\cal D}_\mu (\xi^\nu K^\mu_\nu-\xi^\mu K^\rho_\rho)=
       {\cal D}^{(\mu}  \xi^{\nu)} (K_{\mu\nu}-H_{\mu\nu} K^\rho_\rho).
       \label{eq:momcon}
\end{equation}
In particular, this show that $\xi^\mu$ need only be a Killing vector for the
3-metric $H^{\mu\nu}$ to obtain a strict conservation law on the boundary.

\section{The well-posedenss of the boundary constraint problem}
\label{sec:wellposed}

The conservation conditions (\ref{eq:bmom}) constrain the boundary data for the
nullcone-worldtube problem. I now show that these constraints can be formulated
in terms of a well-posed initial value problem intrinsic to the worldtube.
The traditional $3+1$ decomposition of spacetime used in the Cauchy formalism
is not applicable to the nullcone-worldtube problem since a foliation by
hypersurfaces  with a null normal gives rise to a degenerate $3$-metric.
However, an analogous $2+1$ decomposition can be made on a timelike worldtube.
Let vector and tensor fields intrinsic to the worldtube be denoted by $v^a$, etc. 

Let $t^a$ be an evolution vector field on the worldtube, i.e. the flow of $t^a$
carries an initial spacelike cross section $t=0$  into a $t$-foliation of the
worldtube, with ${\cal L}_{t^a} t =1$. Coordinates $y^A$ of the initial slice
then induce {\em adapted} coordinates $y^a =(t,y^A)$ on the worldtube by
requiring  ${\cal L}_{t^a} y^A=0$. Thus the choice of $t^a$ and initial
coordinates $y^A$ fix the gauge freedom on the worldtube.  

The intrinsic 3-metric $H_{ab}$ of the worldtube has the further  $2+1$
decomposition
\begin{equation}
  H_{ab}=-T_a T_b+ R^2 h_{ab}\, , \quad  H^{ab}=- T^a T^b +R^{-2} h^{ab}.
\end{equation}
where  $T^a$ is the future timelike unit normal to the $t$-foliation and $R^2
h_{ab}$ is the intrinsic 2-metric of the $t=const$ slices, with
$h^{AB}h_{BC}=\delta^A_C$. Here, for later application to the characteristic
problem, We have introduced the surface area factor $R(t,y^A)$,  so that
$\det(h_{AB})=\det(Q_{AB})=Q$, where $Q_{AB}$ is a unit sphere metric.  For
convenience, we chose stereographic coordinates $y^A=(q,p)$ for the unit sphere
metric for which $Q_{AB}dy^A dy^B = Q^{1/2}(dq^2+dp^2)$ with
$Q^{1/2}=4/(1+q^2+p^2)^2$. Since $h^{AB}$ has $(++)$ signature, it can be put in
the matrix form
\begin{equation}
    h^{AB}= Q^{-1/2}\left( 
\begin{array}{cc}
    e^{-2\gamma} \cosh2 \alpha & -\sinh 2 \alpha \\
    -\sinh 2\alpha& e^{2\gamma} \cosh 2\alpha                 
\end{array}
    \right)  ,
\label{eq:gamalph}
\end{equation}
where $\gamma$ and $\alpha$ represent the two degrees of freedom. A specific
choice of polarization dyad associated with this representation is
\begin{equation}
      m^A= Q^{-1/4}\bigg ( e^{-\gamma}(\cosh\alpha-i \sinh\alpha), 
             i e^{\gamma}(\cosh\alpha+i \sinh\alpha)\bigg ) .
\end{equation}

The operator
\begin{equation}
               h^a_b =\delta^a_b + T^a T_b
\end{equation}
projects tensor fields into the tangent space orthogonal to $T^a$. The
relations 
\begin{equation}
      t^a=A T^a + B^a \, , \quad T_a=-A\partial_a t
\end{equation}
are the $2+1$  relations defining the lapse $A$ and shift $B^a$
on the worldtube analogous to the
standard $3+1$ decomposition of the Cauchy problem.  In the adapted coordinates,
$H_{AB}=R^2 h_{AB}$ is the metric of the 2-surfaces of constant $t$
on the worldtube, 
\begin{equation}
  A^2=- H_{tt}+H_{AB}B^A B^B
  \label{eq:laps}
\end{equation}
is the square of the lapse function and 
\begin{equation}
   B^A=R^{-2}h^{AB}H_{tB}
    \label{eq:shif}
\end{equation}
is the shift vector. We have $T_a= -A\partial_a t$, the contravariant
components of the worldtube metric are
\begin{eqnarray}
       H^{tt}&=&-A^{-2} \nonumber \\
        H^{tA}&=&A^{-2}B^A \label{eq:cH} \\
       H^{AB}&=& R^{-2}h^{AB} -A^{-2}B^A B^B. \nonumber
\end{eqnarray}
and we again have the dyad decomposition  $h^{ab}=m^{(a }\bar m^{b)}$
with $m^a=(0,m^A)$. 

The mathematical structure of the conservation conditions (\ref{eq:bmom}) is simplest
to analyze in a worldtube gauge in which the lapse $A=1$ and the shift $B^A=0$. This
corresponds to the introduction of Gaussian normal coordinates on the worldtube, which
is always possible locally.
In this Gaussian gauge,  $H_{tt}=H^{tt}=-1$, $H_{tA}=H^{tA}=0$,  $H^{AB}= R^{-2}h^{AB}$and 
$T_a= -\partial_a t$.

Let ${\cal D}_a$ denote the 3-connection on the worldtube associated with
$H_{ab}$. Then in the Gaussian gauge, the conservation conditions (\ref{eq:bmom}) reduce to
\begin{equation}
    {\cal D}_b(K_a^b- \delta^b_a K)=
     \frac{1}{R^2\sqrt{Q}}\partial_b \bigg(R^2\sqrt{Q}( K^b_a-\delta^b_a K)\bigg )
         +\frac{1}{2} H^{BC}_{,a}(K_{BC}-H_{BC} K)=0
       \label{eq:scon}
\end{equation}
with components
\begin{eqnarray}
   \frac{1}{R^2}\partial_t (h^{BC}K_{BC}) - {\cal D}_B K^B_t
          -\frac{1}{2}H^{BC}_{,t}(K_{BC}-H_{BC}K) &=&0\\
     \frac{1}{R^2} \partial_t(R^2 K_{At}) -{\cal D}_B K^B_A +{\cal D}_A K&=&0.
 \end{eqnarray}
These equations can be re-expressed in terms of the 2-connection $d_A$ on the
slices of the worldtube associated with $h_{AB}$ as 
\begin{eqnarray}
       \partial_t (h^{CD}K_{CD}) - d_A (h^{AB} K_{tB})&=&
          \frac{1}{2}h^{BC}_{,t}(K_{BC} -\frac{1}{2}h_{BC} h^{DE}K_{DE})
                -\frac{1}{R}R_{,t}(h^{BC}K_{BC}-2R^2 K)
                \label{eq:st}\\
     \partial_t( h^{AB} K_{tB})- \frac{1}{2}H^{AB}d_B(h^{CD}K_{CD} )&=& 
        R^2  H^{AB}_{,t}  K_{tB} +h^{BC} d_B (K^A_C - \frac{1}{2}h^A_C K^D_D)
              -h^{AB} d_B K
               \nonumber \\
           &+& \frac{2}{R}R_{,B}(h^{BC}K^A_C-h^{AB}K^C_C) .
            \label{eq:sA}
 \end{eqnarray}

We interpret (\ref{eq:st}) - (\ref{eq:sA}) as a system of equations for $h^{BC}K_{BC}$ 
and $K_{At}$ given their right hand sides. As such, they take the form
of a {\em symmetric hyperbolic} 
system  (cf.~\cite{krlor}). This requires that, up to lower differential terms,
the left hand sides have the form
\begin{equation}
   \partial_t V^{\alpha} -M^{\alpha A}_{\beta} \partial_A  V^\beta
\end{equation}
where, for each  $A$-component,  ${\cal H}_{\gamma \alpha}M^{\alpha A}_{\beta}$
is symmetric in $(\gamma,\beta)$ for some symmetric, positive-definite matrix
${\cal H}_{\gamma \alpha}$. In the present case, we have
$V^\alpha={}^T (V^t,V^q,V^p)=(h^{CD}K_{CD}, h^{qB}K_{Bt},h^{pB}K_{Bt})$.
The component $M^{\alpha q}_{\beta}$ is given  (up to lower order terms) by
\begin{equation}
    M^{\alpha q}_{\beta}=\frac{1}{2} \left( 
\begin{array}{ccc}
    0 &2& 0 \\
    H^{qq}& 0 & 0 \\            
    H^{qp}& 0  &  0  
\end{array}
    \right) .
\end{equation}
Referring to (\ref{eq:gamalph}), the symmetrizer is
\begin{equation}
   {\cal H}_{\gamma \alpha}= \left( 
\begin{array}{ccc}
    (2R^2\cosh  2\alpha)^{-1} & 0& 0 \\
    0 &e^{2\gamma} & \tanh 2\alpha \\            
    0 & \tanh 2\alpha & e^{-2\gamma}   
\end{array}
   \right) .
\end{equation}
The component $M^{\alpha p}_{\beta}$  has the same symmetrizer.

Such symmetric hyperbolic systems have a well-posed Cauchy problem, i.e. there
exists a unique solution which depends continuously on the values of
$h^{BC}K_{BC}$ and $K_{tA}$ on  the initial slice of the worldtube. In addition
to lower order terms in $V^\alpha$, the right hand sides of (\ref{eq:st}) and
(\ref{eq:sA}) depend upon  $K^A_B-(1/2)h^A_B K^C_C$, $K$, $h_{AB}$ and $R$. 
Here  $K^A_B-(1/2)h^A_B K^C_C$ is determined by $m^a m^b K_{ab}$ or, in
tensorial form, by $(h^a_c h_b ^d-\frac{1}{2} h^a_b h^d_c)K^c_d$. Thus, in the
Gaussian gauge, we have established the following {\bf Worldtube Theorem}:

\bigskip

\noindent {\em Given $H_{ab}$, $m^a m^b K_{ab}$ and $K$, the worldtube
constraints constitute a well-posed initial-value problem which determines the
remaining components of the extrinsic curvature $K_{ab}$}.

\bigskip

This theorem extends to any gauge. More generally, the conservation conditions
(\ref{eq:bmom}) take the form
\begin{eqnarray}
      T^b \partial_b (h^d_cK^c_d ) - d_B (h^B_cK^c_d T^d )&=& {\cal S}_t
                        \label{eq:gst}\\
     T^b \partial_b (h_A^cK_c^d T_d ) -\frac{1}{2} d_A (h^d_c K^c_d )&=& {\cal S}_A,
            \label{eq:gsA}
 \end{eqnarray}
where the source terms ${\cal S}_t$ and ${\cal S}_A$ are determined by $H_{ab}$
, $m^a m^b K_{ab}$, $K$ and lower order terms. The system (\ref{eq:gst}) -
(\ref{eq:gsA}) can again be symmetrized.

It is important to note that whether or not $H_{ab}$, $m^a m^b K_{ab}$ and $K$ can be
chosen independently depends upon the choice of gauge conditions and formulation
of Einstein's equations. In the next Section, we discuss this issue in the
context of the Bondi-Sachs formulation.

\section{Conservation laws and the characteristic integration constants}
\label{sec:intconst}

The worldtube theorem established in Sec.~\ref{sec:wellposed} formulates the
boundary constraints as a well-posed initial value problem given the required
source terms. We now explore the implications for the integration constants
necessary to integrate the characteristic hypersurface and evolution equations.
The integration constants  (\ref{eq:intc}) constitute 8 real functions on the
worldtube. The areal radius of the worldtube $R(u,x^A)$ is a 9th function
necessary to determine a unique solution. Thus no more than 9 pieces of
worldtube data can be freely specified. This matches the number of functions
assumed in the worldtube theorem, i.e. $H_{ab}$, $m^a m^b K_{ab}$ and $K$.
However, the boundary constraints introduce 3  relations between the 9 pieces of
worldtube data.

We denote by $y^a=(t,y^A)$ the coordinates intrinsic to a foliation of the
worldtube by topological spheres $t=const$, with angular coordinates $y^A$. The 
propagation of these coordinates along the null geodesics of the  outgoing
$u=const$ null hypersurfaces, which are uniquely determined by this foliation,
induces Bondi-Sachs coordinates $x^\mu=(u,r,x^A)$ in the exterior spacetime with
$x^a=(u,x^A)=y^a$ on the worldtube.  When convenient we will switch from
4-dimensional notation with coordinates $x^\mu$ to 3-dimensional notation with
coordinates $y^a$. We have
\begin{equation}
         \frac{\partial}{\partial y^a} = \frac{\partial}{\partial x^a} +R_{,a}
     \frac{\partial}{\partial r}
         \label{eq:yxR}
\end{equation}
where we continue to denote partial derivatives by commas when it does not lead
to ambiguity. In addition, in this Section, we denote  $\partial / \partial y^a
=\partial_a$.

The metric on the worldtube $r=R(u,x^A)$ which is induced by the Bondi-Sachs
metric (\ref{eq:bmet}) is
\begin{equation}
  H_{ab}dy^a dy^b=-e^{2\beta}{V \over R}dt^2 -2e^{2\beta} R_{,a}dtdy^a
        +R^2h_{AB}(dy^A-U^Adt)(dy^B-U^Bdt).
        \label{eq:bdrymet}
\end{equation}
Here $V(t,y^A)$ is related to the Bondi-Sachs variable $V(u,r,x^A)$ by
$V(t,y^A)=V(t,R(t,y^A),y^A)$. Similarly, $h_{AB}(t,y^A)=h_{AB}(t,R(t,y^A),y^A)$,
etc. We again denote by $d_A$ the 2-connection on the slices of the worldtube
associated with $h_{AB}$. It is important to distinguish between $d_A$ and the
2-dimensional covariant derivative $D_A$  associated with $h_{AB}$ on the 
$(r=const, u=const)$ Bondi-Sachs spheres.  In the tangent space of the
worldtube, we again have the dyad decomposition  $h^{ab}=m^{(a }\bar m^{b)}$
with $m^a=(0,m^A)$. It is also important to distinguish between vectors tangent
to the worldtube and vectors tangent to the $r=const$ Bondi-Sachs coordinate
surfaces.  In the Bondi-Sachs coordinates $x^\mu$, a vector $v^\mu$ tangent to the
worldtube has  components $v^\mu =(v^a,v^r)$ where $v^r=v^aR_{,a}$. Thus $m^\mu
\partial_\mu = m^A\partial_A + m^A R_{,A} \partial_r$. This same procedure
allows us to express $T^a$ and $H^{ab}$ in terms of  Bondi-Sachs components
$T^\mu$ and $H^{\mu\nu}$, e.g $T^\mu \partial_\mu = T^a \partial_a + T^a
R_{,a}\partial_r$. In this regard, it is useful to recall that forms pull back
and vectors (and other contravariant fields) push forward under the embedding
map from the worldtube to the spacetime.

In order to interpret the implication of the boundary constraints for the
nullcone-worldtube problem we now relate the intrinsic metric and extrinsic
curvature of the worldtube to the integration constants (\ref{eq:intc}) for the
hypersurface and evolution equations.  From (\ref{eq:laps}) and (\ref{eq:shif}),
the lapse and shift corresponding to (\ref{eq:bdrymet}) are given in terms of
Bondi-Sachs variables by
\begin{equation}
  A^2= e^{2\beta}( \frac{V}{R}+2R_{,t})+R^2 h_{AB}(B^A+U^A)(B^A-U^A)
            \label{eq:lapse}
 \end{equation}
and 
\begin{equation}
   B^A=-R^{-2} e^{2\beta}h^{AB}R_{,B} -U^A.
   \label{eq:shift}
\end{equation}

The geometric properties associated with the $t$-foliation of the boundary, with
normal $T_a= -A\partial_a t$,  are the 3-acceleration
\begin{equation}
         a^b=T^a {\cal D}_a T^b 
\end{equation}
and the 2-dimensional extrinsic curvature of the $t$-foliation
\begin{equation}
  \kappa_{ab}= h^c_a {\cal D}_c T_b .
\end{equation}

The independent components of $a^b$ and $\kappa_{ab}$ are
\begin{equation}
      m^b a_b= m^B d_B \log A
\end{equation}
\begin{equation}
      h^{ab} \kappa_{ab}= -A^{-1}\bigg( d_C (R^2 B^C)-\partial_t(R^2) \bigg)
\end{equation}
and
\begin{equation}
      m^a  m^b \kappa_{ab}= \frac{R^2}{2A}m^A m^B \partial_t h_{AB}
              -\frac{R^2}{A} m^C m_D d_C B^D .
\end{equation}

The 3-dimensional extrinsic curvature of the worldtube is
$K_{\mu\nu}=H_\mu^\alpha \nabla_\alpha N_\nu$ where the unit outward normal
is given by
\begin{equation}
     N_\mu= \eta \nabla_\mu( r-R).
\end{equation}
The normalization condition expressed in Bondi-Sachs coordinates gives
\begin{equation}
         \eta^{-2}= \frac{V}{R}e^{-2\beta}+2 e^{-2\beta}R_{,u} 
      +2 e^{-2\beta}U^A R_{,A}
               +R^{-2}h^{AB}R_{,A} R_{,B} = e^{-4\beta}A^2,
\end{equation}
so that $\eta=e^{2\beta}A^{-1}$. The outgoing null vector normal to the
$t$-foliation of the worldtube is
$$
    K^\mu=T^\mu +N^\mu .
$$
In Bondi-Sachs coordinates, $K^\mu \partial_\mu$ is proportional to
$\partial_r$. The proportionality constant is determined by the normalization
condition $ K^\mu N_\mu= 1$, which evaluated on the worldtube gives 
\begin{equation}
          K^\mu \partial_\mu = \eta^{-1} \partial_r =e^{-2\beta} A \partial_r.
          \label{eq:Kpartial}
\end{equation}

The computation of $K_{\mu\nu}$ in terms of Bondi-Sachs coordinates can be
simplified by the construction
\begin{eqnarray}
   K_{\mu\nu}&=&H_\mu^\alpha \nabla_\alpha N_\nu
                          = H_{(\mu}^\alpha H_{\nu)}^\beta \nabla_\alpha N_\beta \\
         &=&   H_{(\mu}^\alpha H_{\nu)}^\beta \nabla_\alpha K_\beta
                   -  H_{(\mu}^\alpha H_{\nu)}^\beta {\cal D}_\alpha T_\beta \\
          &=& \frac{1}{2} H_\mu^\alpha H_\nu^\beta {\cal L}_K   g_{\alpha\beta}
                 -    h_\mu^\alpha h_\nu^\beta   {\cal D}_\alpha T_\beta 
                 + h_{(\mu}^\alpha T_{\nu)}T^\beta {\cal D}_\beta T_\alpha
\end{eqnarray}
so that
\begin{equation}
  K_{\mu\nu}= \frac{1}{2}  H_\mu^\alpha H_\nu^\beta {\cal L}_K   g_{\alpha\beta}
           -\kappa_{\mu\nu} + a_{(\mu}T_{\nu)}.
           \label{eq:lieK}
 \end{equation}
  
Expressions for the components of $ {\cal L}_K   g_{\alpha\beta}$ in terms of
the Bond-Sachs variables, which are valid for an arbitrary gauge,  are given in
Appendix~\ref{sec:app}. Collecting the pieces of (\ref{eq:lieK}), the components
of the extrinsic curvature are
\begin{equation}
      h^{ab} K_{ab}= 2ARe^{-2\beta}+A^{-1}\bigg( d_C (R^2 B^C)-\partial_t(R^2) \bigg)
      \label{eq:Kh}
\end{equation}
\begin{equation}
      m^a  m^b K_{ab}=  \frac{1}{2}AR^2e^{-2\beta}m^A m^B h_{AB,r}
       -\frac{R^2}{2A}m^A m^B \partial_t h_{AB}+\frac{R^2}{A} m^C m_D d_C B^D 
        \label{eq:Kmm}
\end{equation}
\begin{equation}
      m^a T ^b K_{ab}   =   -\frac{1}{2}m^AQ_A -m^A R_{,A}(\beta_{,r} -  \frac{1}{2R})  
       + \frac{1}{4}m^A m^B h_{AB,r} \bar m^C R_{,C}-\frac{1}{2}m^C \partial_C \log(e^{-2\beta}A^2) 
        \label{eq:Kmt}
\end{equation}
and
\begin{eqnarray}
    T^a T ^b K_{ab}&=&
          -\frac{1}{A}(\frac {V}{R}+R_{.t}-B^AR_{.A})\beta_{,r} +\frac {V}{2AR^2}
        -\frac {1}{2AR}V_{,r} -\frac {1}{AR^2}e^{2\beta}h^{AB}R_{,B}Q_A \nonumber \\
        &-&\frac{1}{2AR^2}e^{2\beta}h^{AB}_{,r}R_{,A}R_{,B}
        + \frac{1}{AR^3}e^{2\beta}h^{AB}R_{,A}R_{,B} -T^a \partial_a \log(e^{-2\beta}A).
         \label{eq:Ktt}
\end{eqnarray}
\bigskip

The interplay between the boundary constraints on $K_{ab}$ and the
characteristic integration constants is complicated by the choice of gauge and
the choice of free data. We consider two complementary  scenarios which simplify
the discussion of the underlying problems.

\subsection{Waveform extraction}

In the first scenario, the Bondi-Sachs integration constants $(\beta, V,U^A, Q_A, h_{AB},R)$
are obtained from the metric in the neighborhood of the worldtube which is supplied
by the numerical results of a $3+1$ Cauchy evolution. This provides the inner
boundary data for a numerical characteristic evolution on a Penrose compactified
grid, which yields the waveform at  ${\cal I}^+$. This approach,  called
Cauchy-characteristic extraction~\cite{ccm,cce}, is used in numerical relativity
to obtain the waveform at ${\cal I}^+$ without the near field ambiguities
introduced by a finite outer boundary for the Cauchy evolution. In
this application, error in the Bondi-Sachs integration constants results from
numerical error inherent in the Cauchy evolution, from error in computing the Jacobian
from the $3+1$ Cartesian coordinates to the Bond-Sachs coordinates and error from
interpolation onto the worldtube. This feeds into a corresponding error violation of the
worldtube conservation laws, i.e. constraint violation.

Alternatively, the Cauchy data can be used to provide $(H_{ab}, m^a m^b
K_{ab},K)$ on the boundary. The worldtube theorem can then be applied
to determine the remaining components of the extrinsic curvature
$(h^a_b K^b_a, m^a T_b K^b_a) $ using the
well-posed evolution system provided by the conservation laws.
This approach could be used to control the constraint violation in the characteristic
evolution which results from error in the Cauchy data.

The implementation would proceeds as follows.
Since $(A,B^C,R)$ are provided from $H_{ab}$,  the component
$h^a_b K^b_a$, as given in (\ref{eq:Kh}), determines the integration constant
for $\beta$. The shift $B^C$ then determines the integration constant for $U^A$
via (\ref{eq:shift}); and the lapse $A$ then determines the
integration constant for $V$ via (\ref{eq:lapse}).
Next, $m^a T_b K^b_a$ determines the integration constants for $Q^A$ in the following
way. From $m^a m^b K_{ab}$, as given in (\ref{eq:Kmm}), we can determine
$\sigma:=\frac{1}{4}m^A m^B \partial_r h_{AB}$, the {\em optical shear} of the null
hypersurfaces emanating outward from the worldtube. Given $\sigma$, we can
determine $\partial_r \beta$ using the hypersurface equation (\ref{eq:beta}).
The integration constant for $Q^A$ is then determined from
$m^a T_b K^b_a$, as given by (\ref{eq:Kmt}).

We have thus shown that the worldtube theorem leads to the following
{\bf Extraction Corollary}:

\bigskip

\noindent {\em Given $H_{ab}$, $m^a m^b K_{ab}$ and $K$, the worldtube
constraints constitute a well-posed initial-value problem which determine the
the Bondi-Sachs integration constants $(\beta, V,U^A, Q_A, h_{AB},R)$}.

\bigskip

\subsection{The initial-boundary value problem}

In the forgoing waveform extraction scenario,  the nine integration constants
$(\beta, V,U^A, Q_A, h_{AB},R)$, which are required to integrate the Bondi-Sachs
equations were supplied by a Cauchy evolution in a manner consistent with
Einstein's equations. The data consisted of the boundary metric $H_{ab}$,
(6 functions), $m^a m^b K_{ab}$ (2 functions) and $K$ (1 function). The
constraints were then enforced via the worldtube theorem to determine the
remaining components of $K_{ab}$, which in turn supplied the
integration constants.

In an initial-boundary value problem, boundary data consistent with the
constraints must be prescribed {\em apriori}, i.e.  before the evolution is
carried out. Enforcement of the boundary constraints has been the major
difficulty in attempting to show that the various formulations of the gravitational
initial-boundary value problem are well-posed. In this regard, the characteristic 
formulation is no exception. The coupling between the Bondi-Sachs evolution
system and the boundary constraint system is complicated. The details
depend upon the choice of free boundary data and the choice of gauge
conditions adopted on the boundary. We illustrate this problem with
two examples. 

\subsubsection {Constant $R$ boundary data}

Consider first the case in which  the the 5 worldtube integration constants for
$(\beta, U^A,h_{AB})$ are prescribed freely along with a constant value of the
areal radius $R$. We then attempt to prescribe the remaining 3 integration
constants for $(V,Q_A)$ via the 3 boundary constraints (\ref{eq:gst}) and
(\ref{eq:gsA}). For simplicity, assume the boundary data are $\beta=U^A=0$,
$h_{AB}=Q_{AB}$ (where $Q_{AB}$ is the unit sphere metric), along with
$R=const$. The lapse (\ref{eq:lapse}) and shift (\ref{eq:shift}) corresponding
to this data reduce to
\begin{equation}
     A^2 =V/R\, , \quad B^A=0.
\end{equation}
Along with the initial data at $t=0$,
\begin{equation}
       h_{AB}(0,r,x^C)=Q_{AB}\, , \quad V(0,R,x^C) =R-2M \, , \quad  Q_A(0,R,x^C) =0,
\end{equation}
this worldtube data determine a
mass $M$ Schwarzschild spacetime in spherically symmetric Bondi coordinates.

Now, with the same boundary data, let the initial data $h_{AB}(0,r,x^C)$ consist
of a pulse whose support is isolated from the worldtube. This evolves to produce
ingoing radiation so that the spacetime in the neighborhood of the boundary is
no longer Schwarzschild  in the domain of dependence of the initial pulse. 
The boundary constraints for this problem reduce to
\begin{equation}
   \partial_t  K^B_B-\frac{1}{A} d_B(A K^B_t ) =0 
   \label{eq:sconstt}
\end{equation}
and
\begin{equation}
   \partial_t (\frac{1}{A} K_{Bt})-\frac{1}{2} d_B(A K^C_C ) =
   d_C\bigg (A (K^C_B-\frac{1}{2}\delta^C_B K^D_D) \bigg) 
          -d_B(AK) + K^C_Cd_B A ,
   \label{eq:sconstA}
\end{equation}
where the components of the extrinsic curvature are
\begin{equation}
      h^{BC} K_{BC}= 2AR 
\end{equation}
\begin{equation}
      m^C  T^b K_{Cb}   =- m^C   (\frac{1}{2}Q_C+\frac{1}{A}\partial_C A ) 
\end{equation}
\begin{equation}
      m^B  m^C K_{ BC}=  \frac{AR^2}{2}m^B m^C h_{BC,r} 
\end{equation}
and
\begin{equation}
    K=\frac {3A}{2R} +A\beta_{,r}   +\frac {1}{2AR}V_{,r}
         +\frac{1}{A^2}\partial_t A.
\end{equation}
Before the pulse hits the worldtube, $V(u,R,x^A)= R-2M$ and $A^2=1-2M/R$, but
afterward the worldtube values of $h_{BC,r}$ and $Q_A$ become time dependent. 

If these boundary constraints were to constitute a well-posed problem for $
K^B_B$ and $K^B_t$ then they would supply the boundary values of $A$ and $Q_A$
necessary to determine the remaining Bondi-Sachs integration constants. However,
two problems arise from the right hand side of (\ref{eq:sconstA}). First, the
term $K^C_A -\frac{1}{2}\delta^C_A K^D_D$ is determined by the optical shear
$\sigma$ at the boundary. But this shear is not known until the
evolution is carried out and  it depends upon the detailed shape of the initial
radiation pulse. Thus the boundary system (\ref{eq:sconstt}) -
(\ref{eq:sconstA}) is coupled to the evolution system.

The second problem is more serious. For this system the evaluation of the
hypersurface equations on the boundary implies
\begin{equation}
     \beta_{,r} =- \frac{R}{16}h^{AB}_{,r}h_{AB,r}= \frac{R}{2}\sigma \bar \sigma
\label{eq:sbeta}
\end{equation}
and
\begin{equation}
        V_{,r} = 1 +\frac{1}{2}h^{CD}d_C Q_D-\frac{1}{4}h^{CD}Q_{C,r}Q_{D,r}
\label{eq:sv}
\end{equation}
so that
\begin{equation}
    K=\frac {3A}{2R}+ \frac{AR}{2}\sigma \bar \sigma
     +\frac {1}{2AR}( 1 +\frac{1}{2}h^{CD}d_C Q_D
      -\frac{1}{4}h^{CD}Q_{C,r}Q_{D,r})
         +\frac{1}{A^2}\partial_t A.
\end{equation} As a result, in addition to coupling the boundary and evolution
systems, the term $d_B (AK)$ on the right hand side of  (\ref{eq:sconstA})
introduces a $h^{CD}d_B d_C Q_D$ term which alters the principle part
of the boundary system so that it is no longer guaranteed to determine
$A$ and $Q_A$ in a well-posed manner.
 
\subsubsection{ Trace $K$ boundary data}

The problem in the preceding example arises because the boundary data for $R$
and $K$ (the trace of the extrinsic curvature of the boundary) cannot in general
be given independently, as is required for application of the worldtube theorem.
Specification of  $R=const$ boundary data simplifies the boundary system since
(\ref{eq:yxR}) then reduces to $\partial/ \partial y^a = \partial/ \partial
x^a$. However, this simplification is offset by the complicated way in which $K$
changes the principle part of the boundary system and affects the well-posedness
through the $d_B (AK)$ term in (\ref{eq:sconstA}).

The only way this complication can be avoided is to prescribe $K(t,y^A)$ as
explicit boundary data, so that it does not affect the principle part of the
boundary system. In Bondi-Sachs coordinates, $R$ and $K$ 
are related by
\begin{equation}
      K= H^{\mu\nu}\nabla_\mu N_\nu=
     \eta H^{\mu\nu}\partial_\mu    \partial_\nu(r-R)
             - H^{\mu\nu}\Gamma^\rho_{\mu\nu} N_\rho
       =-\eta H^{ab}\partial_a \partial_b R 
             - H^{\mu\nu}\Gamma^\rho_{\mu\nu} N_\rho.
             \label{eq:Rwave}
\end{equation}
Here the timelike nature of the worldtube ensures that $H^{ab}\partial_a
\partial_b$ is a wave operator. Thus, given $K(t,y^A)$, the lapse $A$,
the shift $B^A$, the conformal 2-metric  $h^{ab}$ and the Bondi-Sachs
Christoffel  symbols $H^{\mu\nu}\Gamma^\rho_{\mu\nu} N_\rho$ on the worldtube,
$R$ can be determined  from its initial data by a well-posed
quasilinear wave problem based upon (\ref{eq:Rwave}). This relationship between
the function locating the worldtube and the extrinsic curvature scalar of the worldtube
was first pointed out in the Friedrich-Nagy~\cite{friednag} treatment of the
initial-boundary value problem.

Although this might at first sound like a promising approach, it leads to
serious difficulties. Foremost, if $R_{,A}\ne 0$ then the
boundary constraint (\ref{eq:gsA})
couples the evolution system with the boundary system in a way which makes the
formulation of a well-posed evolution-boundary system appear to be intractable.
This arises from the terms
$$
     -m^A R_{,A}\beta_{,r}+\frac{1}{4} m^A m^B h_{AB,r} \bar m^C R_{,C} 
$$
in the expression (\ref{eq:Kmt}) for $m^a T^b K_{ab}$. Here $\beta_{,r} $ and
$h_{AB,r}$ cannot be determined without knowledge of the evolution.
Consequently,  the boundary constraint governing $m^a T^b K_{ab}$ cannot
be used to determine the Bondi-Sachs integration constant for $Q_A$
independently of the evolution.

One possible way to circumvent this problem would be to pick a gauge for the
boundary in which $R_{,A}=0$, i.e. $R=R(t)$. Since $R$ is a scalar density
defined by the determinant of the 2-metric of the boundary slices,  this can be achieved via
the Jacobian of an appropriately chosen angular transformation. However, this
raises the new problem of how to pick explicit data for $K$ which would be
consistent with $R_{,A}=0$. 

\section{Discussion}

We have shown how the boundary constraints for the Bondi-Sachs equations can be
posed as a symmetric hyperbolic system governing the evolution of certain
components of the extrinsic curvature  of the worldtube, as described by the
worldtube theorem in Sec.~\ref{sec:wellposed}.  In
Sec.~\ref{sec:intconst} we described how these extrinsic curvature components
were related to the integration constants for the Bondi-Sachs system. The
application of the worldtube theorem requires knowledge of the intrinsic metric
of the worldtube and the remaining components of the extrinsic curvature. We
considered two different versions. 

The first application was to waveform extraction. In that case, the data
$(H_{ab}, m^a m^b K_{ab},K)$ necessary to apply the worldtube
theorem are supplied by the numerical results of a $3+1$ Cauchy evolution.
The remaining components of the extrinsic curvature
can then be determined by means of a well-posed initial value problem on the
boundary. The integration constants  $(\beta, V,U^A, Q_A, h_{AB},R)$, for the
Bondi-Sachs equations are then detemined. This
approach can be used to enforce the constraints in the numerical computation of
waveforms at  ${\cal I}^+$ by means of Cauchy-characteristic extraction. 

In the second application, we considered the initial-boundary value problem, for
which boundary data consistent with the constraints must be prescribed  {\em
apriori}, i.e. independent of the evolution. The object was to obtain a
well-posed version of the characteristic initial-boundary value problem.
However, the complicated coupling between the
Bondi-Sachs evolution system and the boundary constraint system prevented any
definitive results. Two choices of free boundary data and boundary gauge
conditions were explored. In both cases, the  Bondi-Sachs choice of
areal coordinate $r$ complicated the analysis. This results from the
way that the coordinates and geometry are mixed, i.e. $r$
cannot be assigned freely on the boundary without specifying its area. It is
possible that other formulations of the characteristic initial-boundary value
problem might be more amenable. Bartnik~\cite{bart} previously explored a
quasi-spherical version, in which the 2-metric $h_{AB}$ is transformed into a
conformally unit sphere form. He found similar complications in trying to
establish well-posedness as in the Bond-Sachs case. A formal computational
algorithm for the evolution-constraint system was possible, but the
well-posedness of the corresponding initial-boundary value problem, which is necessary
for numerical stability, was not clear.

Another possibility is to choose a gauge in which the areal coordinate $r$
is replaced by an affine parameter $\lambda$, so that the affine freedom allows
the specification $\lambda=0$ on the boundary, independently of its geometry.
Rendall~\cite{rend} has shown that such a characteristic initial-boundary value
is well-posed in the double null case where the boundary is also a null
hypersurface. However, Rendall's approach cannot be applied to the corresponding
null-timelike problem. Although the full treatment of the null-timelike problem
lies outside the scope of this paper, it is clear that the wordltube conservations laws
must enter in an essential way.

\appendix
\section{}
\label{sec:app}

We use (\ref{eq:Kpartial}),
\begin{equation}
          K^\mu \partial_\mu =e^{-2\beta} A \partial_r ,
\end{equation}
to simplify the calculation of ${\cal L}_K g_{\alpha\beta}$ in Bondi coordinates. We
have
\begin{equation}
{\cal L}_K   g_{\alpha\beta}=
     e^{-2\beta} A g_{\alpha\beta ,r}
     +  e^{2\beta} A^{-1}   \bigg( 
     K_\alpha \partial_\beta (  e^{-2\beta} A )
     + K_\beta \partial_\alpha (  e^{-2\beta} A ) \bigg ) .
\end{equation}
Thus
\begin{eqnarray}
  H_\mu^\alpha H_\nu^\beta {\cal L}_K   g_{\alpha\beta}&=&
     e^{-2\beta} A H_\mu^\alpha H_\nu^\beta g_{\alpha\beta ,r}
     +  e^{2\beta} A^{-1} H_\mu^\alpha H_\nu^\beta   \bigg( 
     K_\alpha \partial_\beta (  e^{-2\beta} A )
     + K_\beta \partial_\alpha (  e^{-2\beta} A ) \bigg ) \nonumber \\
    &=&
     e^{-2\beta} A H_\mu^\alpha H_\nu^\beta g_{\alpha\beta ,r}
     +  T_\mu {\cal D}_\nu \log  (  e^{-2\beta} A ) 
       +  T_\nu {\cal D}_\mu  \log(  e^{-2\beta} A ) .
\end{eqnarray}
In Bondi-Sachs coordinates, the tangents to the worldtube foliation have
components $m^\mu=(0,m^B R_{,B}, m^A)$ and
$T^\mu=A^{-1}(1, R_{,t} -B^C R_{,C}, -B^A)$. This leads to the components 
\begin{equation}
      m^\mu \bar m^\nu H_\mu^\alpha H_\nu^\beta {\cal L}_K   g_{\alpha\beta} =
            4ARe^{-2\beta} 
\end{equation}
\begin{equation}
      m^\mu m^\nu H_\mu^\alpha H_\nu^\beta {\cal L}_K   g_{\alpha\beta} =
            AR^2e^{-2\beta}m^A m^B h_{AB,r} 
\end{equation}
\begin{eqnarray}
      m^\mu T ^\nu H_\mu^\alpha H_\nu^\beta {\cal L}_K   g_{\alpha\beta} &=&
          -m^A Q_A-2M^A R_{,A}\beta_{,r}-  e^{-2\beta}m^B  g_{AB,r} (U^A+B^A)
      -m^A d_A \log(e^{-2\beta}A)
           \nonumber\\
         &=&   -m^AQ_A -2m^A R_{,A}(\beta_{,r} -  \frac{1}{R})  
       + \frac{1}{2}m^A m^B h_{AB,r} \bar m^C R_{,C}-m^A \partial_A \log(e^{-2\beta}A)  
\end{eqnarray}
and
\begin{eqnarray}
      T^\mu T ^\nu H_\mu^\alpha H_\nu^\beta {\cal L}_K   g_{\alpha\beta} &=&
         -\frac{2}{A}(\frac {V}{R}+2R_{.t}-2B^AR_{.A})\beta_{,r}
        +\frac {V}{AR^2}
        -\frac {1}{AR}V_{,r} -\frac {2}{A}(U^A+B^A)Q_A \nonumber \\
        &+&\frac{1}{A}e^{-2\beta}g_{AB,r}(U^A+B^A)(U^C+B^C)
        -2T^a \partial_a\log(e^{-2\beta}A)
           \nonumber\\
         &=& -\frac{2}{A}(\frac {V}{R}+2R_{.t}-2B^AR_{.A})\beta_{,r} 
         +\frac {V}{AR^2}
        -\frac {1}{AR}V_{,r} -\frac {2}{AR^2}e^{2\beta}h^{AB}R_{,B}Q_A \nonumber \\
        &-&\frac{1}{AR^2}e^{2\beta}h^{AB}_{,r}R_{,A}R_{,B}
        + \frac{2}{AR^3}e^{2\beta}h^{AB}R_{,A}R_{,B} 
               -2T^a \partial_a\log(e^{-2\beta}A).
\end{eqnarray}

\begin{acknowledgments}

This research was supported by NSF grant PHY-0854623 to the University of
Pittsburgh. My thanks again to Josh for his friendship and guidance.

\end{acknowledgments}

\end{document}